\documentclass[numsec , webpdf, modern,large,namedate]{oup-authoring-template}

\graphicspath{{Fig/}}

\theoremstyle{thmstyleone}%
\theoremstyle{thmstyletwo}%
\theoremstyle{thmstylethree}%

\begin{document}

\firstpage{1}

\title[VoDEx: time annotation and management of volumetric functional imaging data]{VoDEx: a Python library for time annotation and management of volumetric functional imaging data}

\author[1,3,$\ast$]{Anna Nadtochiy\ORCID{0000-0002-5520-2289}}
\author[2,3]{Peter Luu\ORCID{0000-0002-5319-9848}}
\author[1,2,3]{Scott E. Fraser\ORCID{0000-0002-5377-0223}}
\author[2,3,$\ast$]{Thai V. Truong\ORCID{0000-0001-6475-8935}}

\authormark{Nadtochiy et al.}

\address[1]{\orgdiv{Department of Quantitative and Computational Biology}, \orgname{University of Southern California}, \orgaddress{\street{Los Angeles}, \state{CA}, \postcode{90089}, \country{USA}}}
\address[2]{\orgdiv{Department of Biological Sciences, Division of Molecular and Computational Biology}, \orgname{University of Southern California}, \orgaddress{\street{Los Angeles}, \state{CA}, \postcode{90089}, \country{USA}}}
\address[3]{\orgdiv{Translational Imaging Center}, \orgname{University of Southern California}, \orgaddress{\street{Los Angeles}, \state{CA}, \postcode{90089}, \country{USA}}}

\corresp[$\ast$]{Corresponding authors. \href{nadtochi@usc.edu}{nadtochi@usc.edu}, \href{tvtruong@usc.edu}{tvtruong@usc.edu}}


\abstract{
\textbf{Summary:}  In functional imaging studies, accurately synchronizing the time course of experimental manipulations and stimulus presentations with resulting imaging data is crucial for analysis. Current software tools lack such functionality, requiring manual processing of the experimental and imaging data, which is error-prone and potentially non-reproducible. We present VoDEx, an open-source Python library that streamlines the data management and analysis of functional imaging data. VoDEx synchronizes the experimental timeline and events (eg. presented stimuli, recorded behavior) with imaging data. VoDEx provides tools for logging and storing the timeline annotation, and enables retrieval of imaging data based on specific time-based and manipulation-based  experimental conditions. \\
\textbf{Availability and Implementation:} VoDEx is an open-source Python library and can be installed via the "pip install" command. It is released under a BSD license, and its source code is publicly accessible on GitHub (\href{https://github.com/LemonJust/vodex}{https://github.com/LemonJust/vodex}). A graphical interface is available as a  napari-vodex plugin, which can be installed through the napari plugins menu or using "pip install." The source code for the napari plugin is available on GitHub (\href{https://github.com/LemonJust/napari-vodex}{https://github.com/LemonJust/napari-vodex}).\newline
\textbf{Contact:} \href{nadtochi@usc.edu}{nadtochi@usc.edu},\href{tvtruong@usc.edu}{tvtruong@usc.edu}
}

\maketitle
\section{Introduction}\label{sec1}
Volumetric functional imaging is widely used in neuroscience studies for recording brain activity in parallel with behavioral and physiological data of organisms \citep{Kim2022}. Such studies often have complex experimental designs, aiming to characterize neural responses to experimental tasks and/or detect differences in brain activity patterns among various stimuli, behaviors, and physiological states.

Accurate analysis of functional imaging data requires accurate annotations of the time course of the experiment and synchronization of the time annotations with the imaging data. Although both commercial and open-source tools are available for designing experiments and tracking behavior \citep{Mathis2018, Peirce2019, Lopes2021, Gabriel2022, Akam2022}, they lack the ability to directly link this information to the imaging data. 

As a result, processing and time-annotating data are often performed manually. Manual data management and processing is not scalable, is hard to document, and is prone to human error, risking potential mistakes that are difficult to detect and correct.

Linking time annotations to volumetric imaging data presents an additional challenge, as volumes are acquired as a series of optical sections:  a series of 2D images are taken sequentially at different depths inside the sample, which are then assembled into a 3D dataset. In order to correctly interpret the volumetric data, it is crucial to track not only the correspondence between image frames and experimental conditions, but also track the exact location of these frames within a volume. Such complexity of time annotation, combined with the unprecedented amount of data produced by functional imaging experiments, makes manual data handling tedious and unreliable. 

To address these challenges, we introduce VoDEx, Volumetric Data and Experiment manager, an open-source  Python library for synchronizing time annotations and volumetric information with imaging data. VoDEx is distributed under a BSD 3-clause License, making it freely accessible for academic, commercial, and personal use. VoDEx integrates the information about individual image frames, volumes, and experimental conditions and allows the retrieval of sub-portions of the 3D-time series datasets based on any of these identifiers without the need to load the entire dataset into memory. It logs all information related to the experiment into an SQLite database, enabling later data verification and sharing in accordance with the FAIR (Findable, Accessible, Interoperable, and Reusable) principles \citep{Wilkinson2016, Dempsey2022}. It is implemented both as a napari \citep{napari} plugin for interactive use with a GUI and as an open-source  Python package.  Python's rich ecosystem of libraries, such as NumPy \citep{harris2020array}, SciPy \citep{2020SciPy-NMeth}, and scikit-image \citep{van2014scikit}, makes VoDEx a useful tool for image analysis and allows for integration into a wide range of analysis pipelines.
\section{Implementation}\label{sec2}
VoDEx contains classes that assist in the creation, organization, and storage of information related to image acquisition and time annotation, allowing for the search and retrieval of image data based on specific conditions. This functionality is split into five modules: core, annotation, dbmethods, experiment, and loaders. 
\begin{itemize}
    \item The core module provides the basic functionality for retrieving image data information. 
    \item The annotation module handles the construction, validation, and storage of time annotation. For cyclic events, VoDEx keeps track of cycle iterations, which is important in behavioral experiments where the subject might become habituated to the repeated stimulus or learn over the course of the experiment. 
    \item The dbmethods module abstracts the SQL calls, providing an easy-to-use interface to populate the SQLite database in which VoDEx stores information, and to query the database. 
    \item The loaders module contains classes designed to load image data from specific file types, with current support for TIFF, and allows for easy addition of support for other file formats. 
    \item The experiment module contains the Experiment class, connecting all the functionalities of the VoDEx package and serving as the main point of entry for interacting with the package.
\end{itemize}

\section{Usage}\label{sec3}
\begin{figure*}[!t]%
\centering
\includegraphics[width=\textwidth]{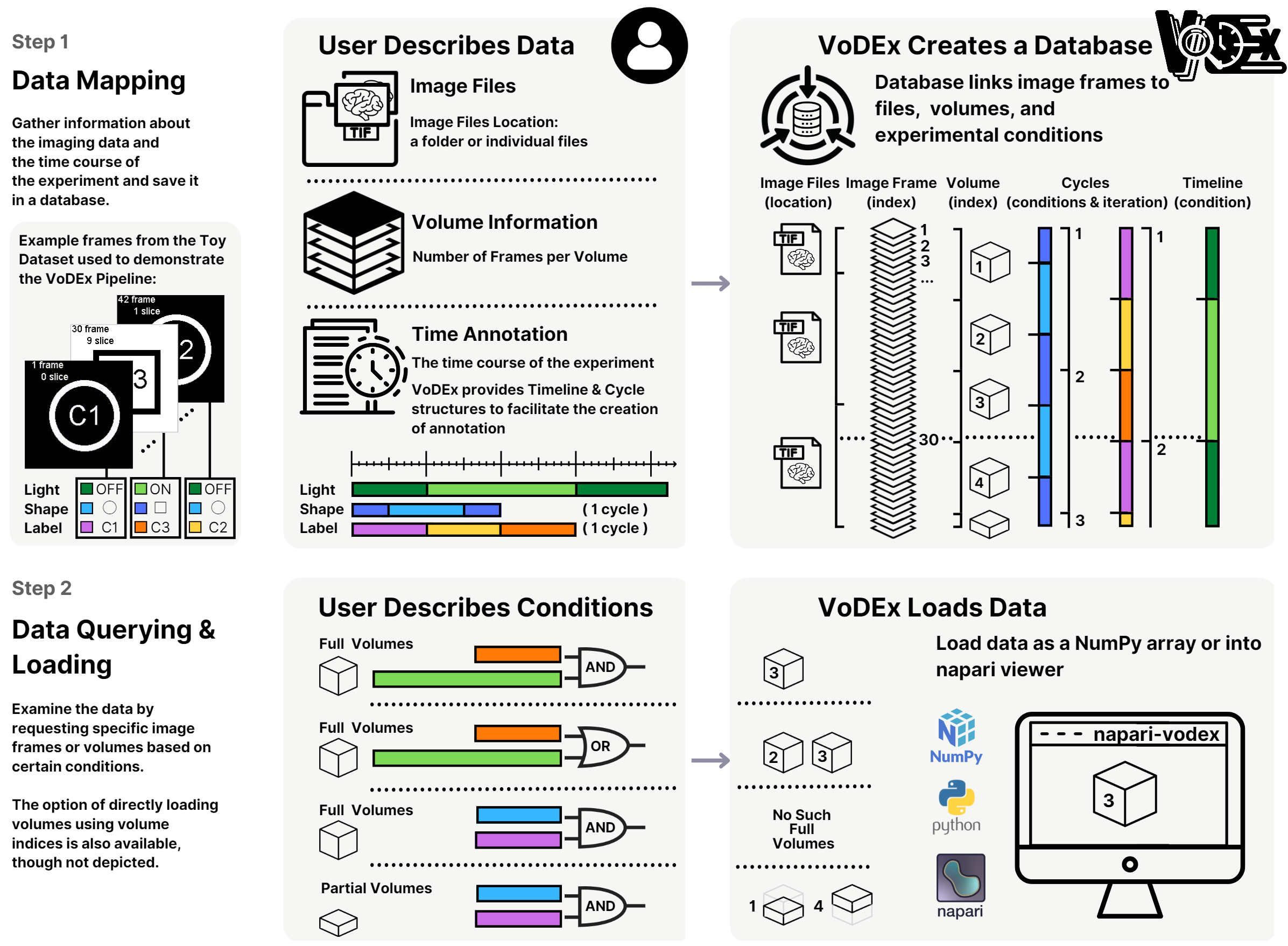}
\caption{An illustration of the VoDEx pipeline applied to a Toy Dataset. The Toy Dataset is available via the project website at \href{https://lemonjust.github.io/vodex/}{https://lemonjust.github.io/vodex/}. The dataset comprises 42 image frames, divided into three TIFF files. Each volume consists of 10 frames, resulting in 4 full volumes, with 2 additional frames at the end of the recording. 
In the dataset, three conditions are tracked: light,  label, and shape. The background of the frame indicates the light (on/off), the middle of the screen shows the label (c1, c2, c3), and the screen displays either a circle or a square.\newline\newline
In Step 1, the user inputs information about the imaging data and the experimental conditions into VoDEx, either through a Python script or a graphical user interface in the napari plugin. Specifically, the user provides the information for the whole recording to encode light, but only one cycle iteration of the label and shape. Note that the shape conditions switch in the middle of a volume. VoDEx then integrates this information, automatically determining the number of frames in each image file, estimating which frames correspond to which volumes, and repeating the provided cycles to cover the duration of the entire recording while keeping track of both conditions and cycle iterations. The information is stored in a database created by VoDEx. For instance, frame number 30 is stored as a 4th frame inside the 3rd TIFF file. It represents the last slice ( 9th slice, zero-based indexing) in the 3rd volume. This frame features a square shape, with the label "c3" and the light "on". It was recorded during the second iteration of the shape cycle and the first iteration of the label cycle.\newline\newline
In Step 2, the user can search and access imaging data based on mapped experimental conditions. Either a Python script or the graphical user interface in the napari plugin can be used. The conditions can be combined with "and" or "or" logic. For instance, if the user wants to view the volumes where the label is c3 and the light is on, VoDEx returns volume index 3. To view the volumes where the label is c3 or the light is on, VoDEx returns both volumes 2 and 3. When searching for full volumes, all frames within the volume must match the conditions. If the conditions do not fully cover a volume, such as when one of the conditions combined with "and" is a shape, working with individual frames is possible. VoDEx also provides options for loading the volumes as numpy arrays for further processing in Python or directly into the napari viewer through the napari plugin.}\label{fig1}
\end{figure*}

The VoDEx pipeline consists of two steps: data mapping and data querying. An illustration of the VoDEx pipeline applied to a Toy Dataset is shown in Fig.\ref{fig1} (Toy Dataset is available via the project website at \href{https://lemonjust.github.io/vodex/data/}{https://lemonjust.github.io/vodex/data/}). During the mapping step, VoDEx creates a mapping between image frames, their location within associated files, the image volumes they correspond to, experimental conditions, and cycle iterations. This information is saved to a database, allowing users to save the experiment description for sharing or to return to it later. In the second step, users can investigate the data by querying the database for image frames or volumes that were recorded during specific combinations of the mapped conditions. The querying process does not require prior knowledge of database query syntax and is conducted via methods provided by VoDEx. When requesting frames, VoDEx returns the indices of all the image frames matching the request. When requesting volumes, VoDEx selects those frames that constitute full volumes and returns the corresponding volume indices. The indexing of frames and volumes start from the beginning of the recording. The image data can thus be loaded based on these indices as a 3D or a 4D numpy array for frames or volumes, respectively.

The package offers a  graphical user interface (GUI) through its integration with napari via the napari-vodex plugin, enabling users to easily navigate the full VoDEx pipeline and to load selected volumes directly into the napari viewer. The two steps of the pipeline can be performed independently using either the script or the GUI. This permits users to annotate the data using the GUI, and to then switch to a script to query the data; alternately, users can perform data annotation in a script, load this annotation into a GUI, and then interactively explore the data in napari. Comprehensive documentation of VoDEx, including examples, tutorials, and toy datasets, are provided at the project website \href{https://lemonjust.github.io/vodex/}{https://lemonjust.github.io/vodex/}. The documentation is continuously updated through the use of mkdocs and mkdocstrings packages, as well as GitHub actions, ensuring that any changes to the API are promptly reflected in the documentation. The package has undergone rigorous testing, achieving 100$\%$ test coverage, through the use of pytest.

\section{Application}\label{sec4}
To demonstrate the capabilities of VoDEx, we present its application to the study of numerosity estimation in zebrafish larvae, where it played a key role in the processing of whole-brain functional imaging data acquired using light-sheet fluorescence microscopy. The numerosity task is designed to explore the Approximate Number System (ANS) of the zebrafish.  Zebrafish larvae were presented with a series of visual stimuli in a pseudo-random order with variable timing. The numerosity stimuli included a blank screen and screens with one to five dots in multiple geometric patterns, which are commonly used to control for potential confounding effects \citep{Piazza2004, Gebuis2016, Pennock2021}.VoDEx facilitated the interpretation of the responses to the stimuli, which required accurate annotation and tracking of various stimulus patterns. VoDEx efficiently managed the sets of visual stimulus patterns and enabled the processing of the large volumetric imaging datasets on a standard computer. The implementation was carried out in Jupyter notebooks, as well as in a custom Python package specifically designed for this study, showcasing the versatility of integrating VoDEx into a comprehensive analysis pipeline. Our preliminary results have been published in Frontiers in Neuroanatomy \citep{Messina2022}, and full analysis pipelines for different numerosity stimuli combinations are available as sets of Jupyter notebooks at \href{https://github.com/LemonJust/numan}{github.com/LemonJust/numan} under notebooks/individual\_datasets.

VoDEx is particularly useful for experiments where the time annotation is derived from experimentally measured events, such as an organism's behavior or physiological data. Unlike well-controlled stimuli sequences that align with the rate of volumetric image collection, transitions between behaviors in such experiments can occur at various times during the recording of individual volumes. VoDEx provides an easy way to detect and manage these types of events.

\section{Conclusion}\label{sec5}
VoDEx is an open-source Python library that streamlines the management and analysis of volumetric functional imaging data. The library offers a suite of tools for creating, organizing, and storing information pertaining to image acquisition and time annotation. Additionally, it allows for the retrieval of image data based on specific experimental conditions, enabling researchers to access and analyze the data easily. VoDEx is available as both a standalone Python package and a napari plugin, providing a user-friendly solution for processing volumetric functional imaging data, even for researchers without extensive programming experience. The library is designed to be flexible and easily integrated with other Python packages, which expands its capabilities for advanced use cases.\newline\newline\noindent
Conflict of interest: None declared.

\section*{Funding}
This work is supported in part by funds from the Human Frontier Science Program (RPG0008/2017) and the National Institutes of Health (1U01NS122082-01, 1U01NS126562-01). 

\section*{Author contributions statement}
A.N. conceived the idea for the software, implemented the software, created documentation, and wrote the main body of the paper. P.L. conducted biological experiments for the study of numerosity estimation in zebrafish. SEF and TVT supervised the work. All authors: reviewed and refined the manuscript and approved the submitted version.

\bibliographystyle{abbrvnat}
\bibliography{reference}

\end{document}